# Analysis of a Mathematical Model of Apoptosis: Individual Differences and Malfunction in Programmed Cell Death

**Elife Zerrin Bagci[1,2,3]*, Sercan Murat Sen[2], Mehmet Cihan Camurdan[2]**

**1** Polymer Research Center, Bogazici University, Istanbul, Turkey, **2** Chemical Engineering Department, Bogazici University, Istanbul, Turkey, **3** Department of Biology, Namik Kemal University, Tekirdag, Turkey

* Corresponding author:

Elife Zerrin Bagci

e-mail: ebagci@nku.edu.tr

# ABSTRACT

Apoptosis is an important area of research because of its role in keeping a mature multicellular organism's number of cells constant hence, ensuring that the organism does not have cell accumulation that may transform into cancer with additional hallmarks. Firstly, we have carried out sensitivity analysis on an existing mitochondria-dependent mathematical apoptosis model to find out which parameters have a role in causing monostable cell survival i.e., malfunction in apoptosis. We have then generated three healthy cell models by changing these sensitive parameters while preserving bistability i.e., healthy functioning. For each healthy cell, we varied the proapoptotic production rates, which were found to be among the most sensitive parameters, to yield cells that have malfunctioning apoptosis. We simulated caspase-3 activation, by numerically integrating the governing ordinary differential equations of a mitochondria-dependent apoptosis model, in a hypothetical malfunctioning cell which is treated by four potential treatments, namely: (i) proteasome inhibitor treatment, (ii) Bcl-2 inhibitor treatment, (iii) IAP inhibitor treatment, (iv) Bid-like synthetic peptides treatment. The simulations of the present model suggest that proteasome inhibitor treatment is the most effective treatment though it may have severe side effects. For this treatment, we observed that the amount of proteasome inhibitor needed for caspase-3 activation may be different for cells in individuals with a different proapoptotic protein deficiency. We also observed that caspase-3 can be activated by Bcl-2 inhibitor treatment only in those hypothetical malfunctioning cells with Bax deficiency but not in others. These support the view that molecular heterogeneity in individuals may be an important factor in determining the individuals' positive or negative responses to treatments.

# Introduction

Chemotherapy resistance is an important problem in cancer treatments. A specific cancer chemotherapy that is used on patients whose tumors have similar histopathology may have very different responses. Oncologists suspect that subsets of patients that respond positively to a chemotherapy are hidden in larger groups of resistant cases. Genetic and molecular heterogeneity may be the cause of those subsets.[1] The presence or absence of, or more specifically, the quantity of a biological molecule e.g., DNA, RNA, protein and other metabolites which indicate whether the individual is healthy or has a disease is a biomarker. Therefore, assessment of biomarkers can unravel this genetic and molecular heterogeneity and may be utilized to determine the type and the intensity of chemotherapy method to treat a patient. Mathematical modeling and computations may facilitate the decision of the chemotherapy method to be used because just experience may not be enough because of the complex nature of cancer.

The beneficial effects of chemotherapy drugs can be mitotic catastrophe, apoptosis or prolonged cell cycle arrest. Hence, defects in apoptotic mechanisms may be a reason for chemotherapy resistance. The mitochondria-dependent apoptotic pathway is the major apoptotic pathway which is utilized by chemotherapeutic drugs.[2] Depending on which tissue they belong to, there are two types of cells for apoptosis: Type I and Type II. If the apoptotic signaling pathway bypasses mitochondria then these are Type I cells and if not, these are called Type II cells.

A mathematical model for mitochondria-dependent apoptosis, in which bistability emerges as healthy functioning of the Type II cells was proposed by Bagci et al.[3] In this model, the extracellular apoptotic stimulus (Fas Ligand) results in cytochrome *c* (cyt *c*) release from the mitochondria and caspase-3 activation which is the executioner enzyme for apoptosis. For simplicity, the model excluded the reactions before caspase-8 formation. The detailed descriptions of the model can be found in Model and Methods section of the study by Bagci et al.[3]

The biochemical mechanism of apoptosis is studied extensively because of the importance of keeping the number of cells in the mature organism balanced in response to pro- or anti-apoptotic stimuli.[4] In healthy tissues, cell number stays constant when the rate of cell proliferation is equal to the rate of cell death. However, in malfunctioning apoptosis, the rate of cell proliferation can be higher (lower) than the rate of cell death and the number of cells increases (decreases). The total number of cells in a tissue increases in tumorigenesis whereas it decreases in neurodegenerative disorders (e.g. Parkinson's disease and Alzheimer's disease).

In this study, by healthy cells, we mean the cells without apoptosis malfunction prior to any treatment. Healthy cells are converted to cells with apoptosis malfunction by decreasing the proapoptotic protein production rates in the model. We call these cells "hypothetical malfunctioning cells". However, these hypothetical malfunctioning cells can not represent all the tumor cells as some of the tumor may have apoptosis rates that are considerably higher than that of normal cell.[5,6] The hypothetical malfunctioning cells are then treated *in silico* by four different treatment methods. The cells are predicted to be resistant to treatment if they do not undergo apoptosis. On the other hand, they are predicted to be sensitive to the treatment if apoptosis is induced.

After pioneering studies by Fussenegger et al.[7] and Eissing et al.,[8] many apoptosis models have been published (see review by Salvioli et al.[9] and other references[3, 10-28]), however, none of these studies focused on resistance to treatment (i.e., despite treatment, the cells survive due to lack of apoptosis). In treatments that target apoptotic pathways, drugs affecting extracellularly either activate Fas or another death signaling receptor. On the other hand, drugs affecting intracellularly upregulate proapoptotic proteins and/or downregulate antiapoptotic proteins. In this study, we focused on intracellular affecting treatments. The results herein suggest that the type of potential treatment and the identity of the deficient proapoptotic protein determine whether apoptosis will be induced in a hypothetical malfunctioning cell. Therefore, the results suggest that the reason of different outcomes of a treatment in different people may be genetic variations in their cells that can be observed through their possible biomarkers for apoptosis namely, production rates of proapoptotic proteins[4].

The previous simulations by Bagci et al.[3] predicted a pathological state in which cells will exhibit a monostable cell survival if the degradation rate constant (expression rate constant) of the proapoptotic protein Bax is above (below) a threshold value. On the other hand, with suitable values of rate constants, the model predicts bistability with a suitable threshold of apoptotic stimulus for apoptosis. We used a mathematical model that was originally proposed by Bagci et al.[3] to shed light on resistance to treatments because this model successfully predicts the correct functioning (bistable, healthy cell) and malfunctioning (monostable, unhealthy cell) of apoptosis. We have also used a modified version of this model in our study. Herein, it is assumed that all pathways other than those related to apoptosis remain unchanged, and therefore a change in apoptosis will lead to a change in homeostasis of the number of cells. In this article, we first present the sensitivity analysis to determine the most sensitive parameters to caspase-3 activation. Then, we summarize the results of the sensitivity analysis performed on the model parameters. This is followed by presenting the simulations of four hypothetical potential treatments i.e., the proteasome inhibitor treatment, Bcl-2 and IAP inhibitors treatment and Bid-like synthetic peptides treatment of which proteasome inhibitor treatment is predicted to be the most effective. The aims of the *in silico* experiments were (i) to gain insights for the role of molecular heterogeneity in resistance to treatment for malfunction in apoptosis and (ii) to check if the underlying reaction mechanisms should be modified and/or new reactions should be added into the pathway and (iii) to check if the parameter values used in the existing model should be known within a narrow range (i.e., sensitive) so that the resulting model and parameters could be used in guiding treatments. The results support the view that molecular heterogeneity among individuals may be a reason for varied responses to treatments. As for the second and third aims, we have found the reactions which are important and the parameters which should be known within a narrow range for the present model. We have also compared the experimental methods of Kim et al.[29] with the theoretical predictions of our modified model. We also compared the predictions of our model with the experimental results compiled by di Pietro et al.[30] from several databases. It should be emphasized that our focus in this study is on malfunction in apoptosis and resistance to treatments for it and not the much more complicated problem, chemotherapy resistance to cancer.

## Methods

In the model proposed in the reference[3], the rate constants ensure bistability in apoptotic response where initial concentrations smaller than threshold values for caspases lead to cell survival and higher initial concentrations lead to apoptosis. The chemical reactions and physical interactions of the proteins involved in the apoptotic pathways of this model are presented in Supplementary Materials Figure S1 and Table S1 for easy reference. As mentioned in the previous section, sensitivity analysis is performed to determine the parameters that may have a role in malfunctioning apoptosis and resistance to treatment. Out of those parameters that are found to be sensitive (ten of them) (See Table 1), we have varied arbitrarily chosen four of them among the benchmark parameter values given in reference[3] to determine whether the system is robust to parameter variations. For this purpose we obtained two additional parameter sets each representing cells belonging to a healthy individual (Table 2). The benchmark parameters are varied so as to preserve the bistable character of the system. Therefore, each parameter set is bistable in response to apoptotic stimulus and hence represents cells of healthy individuals. We then used these three parameter sets to simulate four hypothetical potential treatments. We also generated six malfunctioning point cells around the nominal malfunctioning point cells in parameter set 1 and 3 to evaluate the statistical significance of the results for the two parameter sets. The new parameter sets around set 1 and 3 are presented in Table 4.

When mass action kinetics is applied, the reactions listed in Table S1 lead to rate equations in the form of nonlinear ordinary differential equations.[3] Herein, we solved these equations numerically using the software XPPAUT developed by G. Bard Ermentrout[31] for the simulation of the potential treatments.

*Modified version of mitochondria-dependent apoptosis model:*

The following degradation of caspase-3 reaction is omitted from existing mitochondria-dependent apoptosis model:

<p style="text-align:center">caspase-3 → aminoacids</p>

The following reactions[32] are added to the existing mitochondria-dependent apoptosis model:

<p style="text-align:center">caspase-3·IAP → IAP + caspase-3$_{ubuiquitinated}$</p>

<p style="text-align:center">caspase-3 $_{ubuiquitinated}$ → aminoacids</p>

## Results and discussion

Firstly, some preliminary results on nullclines, and phase plane for bistable and monostable dynamical system are discussed for easy reference as they are profusely used in the later subsections. Figure 1A illustrates the phase plane of a bistable apoptosis model. There are three steady states, denoted by (i), (ii) and (iii) which have zero, high and intermediate levels of executioner caspase (caspase-3) concentration. The steady state (i) is the cell survival state (zero executioner caspase concentration), (ii) is the possible apoptotic state and (iii) is the cell fate decision point. The steady states (i) and (ii) are stable because a small perturbation away from them eventually disappears. On the other hand, the steady state (iii) is unstable because a small perturbation away from it grows. The thick curves are the nullclines[33] and their intersections are the steady state points.

The two stable equilibrium points may bifurcate to one stable equilibrium point (i.e., become monostable) if a system parameter (e.g., degradation rate of caspase-3) is changed. The phase plane in Figure 1B is an example of this case; where the monostable state is cell survival.

We first present the sensitivity of steady state caspase-3 concentration to the parameters of the model by Bagci et al.[3] and determine those that may induce a malfunction in apoptosis when perturbed from their nominal values. The model is presented schematically in Figure S1 and the list of chemical reactions and physical interactions are listed in Table S1. For parameter values, see the reference[3]. Then, we present outcomes for four hypothetical treatments that are obtained by simulating the model for three different parameter sets each representing a healthy cell. Finally, we present the sensitivity results and proteasome inhibitor treatment outcomes for the modified model (corrected for IAP – caspase-3 interactions), and also compare model predictions with experiments that addressed the effect of Bcl-2 and IAP inhibitors in chondrosarcoma cells[29] and with experimental results compiled by di Pietro et al.[30] from databases.

**Sensitivity analysis of mitochondria-dependent apoptosis model**

To evaluate the sensitivity of caspase-3 concentration to the parameters, we used a different and a much simpler method than the one used by Shoemaker and Doyle[34] who have carried out sensitivity analysis on the parameters in the same model using tools from control engineering. However, the results of the two different approaches are in good agreement. Hereafter, the nominal values of the parameters are understood to be those given in the reference[3] and the steady-state concentration of caspase-3 for these parameters corresponds to 5.4 nM. For sensitivity analysis we increased and decreased the values of the parameters by 100-fold and then checked how much the steady-state value of [caspase-3] is changed (Table 1). The following parameters are found to be sensitive that affect the final steady-state concentration of caspase-3: p53 concentration ([p53]), production rate constants of pro-apoptotic proteins, Apaf-1 ($\Omega_{Apaf1}$), procaspase-3 ($\Omega_{proc3}$), procaspase-9 ($\Omega_{proc9}$), Bid ($\Omega_{Bid}$), Bax ($\Omega_{Bax}$), mitochondrial cyt $c$ ($\Omega_{cytcmito}$), anti-apoptotic proteins Bcl-2 ($\Omega_{Bcl2}$), IAP ($\Omega_{IAP}$), and degradation rate constant for all the proteins ($k_d$). The same results were also observed by Shoemaker and Doyle[34]. They also found that steady-state concentration of caspase-3 is not sensitive to most of the parameters.

Sensitivity analysis revealed not only the sensitive parameters but also whether an increase or decrease in these parameters may lead to malfunction in apoptosis. It is found that when the production rate constants of proapoptotic proteins are low, and production rate constants of antiapoptotic proteins and degradation rate constant of all proteins are high, the model predicts monostability with the cell survival state being the only stable state. Therefore, the sensitivity analysis helped us to create cells with malfunctioning apoptosis. These hypothetical malfunctioning cells are then treated *in silico* by four treatment types.

**Hypothetical treatments**

The roles of the sensitive parameters on malfunction in apoptosis and resistance to treatments in a hypothetical cell were then investigated. The simulations were carried out for three parameter sets representing three healthy cells which may belong to three different individuals (since each parameter set results in bistability) to test the robustness of the theoretical outcomes of the treatments. Parameter set 1 was taken as the parameter values used in the reference[3]. Parameter set 2 was obtained by changing the numerical values of production rates of IAP ($\Omega_{IAP}$), procaspase-9 ($\Omega_{proc9}$), procaspase-3 ($\Omega_{proc3}$) of set 1. Finally, parameter set 3 was obtained by changing the numerical values of production rates of IAP ($\Omega_{IAP}$), procaspase-3 ($\Omega_{proc3}$) and mitochondrial cyt $c$ ($\Omega_{cytcmito}$) of set 1. These three parameter sets are given in Table 2. We note that the numerical values of parameters in parameter set 1 are close to those in set 2 but considerably different than those in set 3. The true values of these sensitive parameters are only known within a wide range. Even if we only take upper and lower values of these ten sensitive parameters (i.e., 2 levels), the number of *in silico* experiments that has to be carried out is a very large value 1024 (=$2^{10}$) to obtain all main and interaction effects.[35] Therefore, we took only three of them to generate our healthy cells which may belong to three different individuals and yet found qualitative and quantitative differences. We note that the gene expression levels of caspase-3, caspase-9 and cyt $c$ are reported to vary naturally in human individuals in a database that also reports certain polymorphisms in the genes.[36]

In order to generate cells with malfunctioning apoptosis from each parameter set, the numerical value of one of the sensitive parameters was changed and then checked whether this cell had malfunction in apoptosis. If the result was on the affirmative, then we applied four potential treatments, one at a time to see if apoptosis can be achieved in this cell. To this end, the production rates of proapoptotic proteins were reduced to 1 % and 15 % of their nominal values. These proapoptotic proteins were Apaf-1, procaspase-3, procaspase-9, Bid, Bax and mitochondrial cyt $c$. The criterion for malfunction in apoptosis in those phenotypes was either monostability (cell survival i.e., caspase-3 concentration is zero as shown in Figure 1B) in responding to apoptotic stimulus or bistability with caspase-3 concentration not reaching a predetermined threshold value of 1 nM which is only to be understood relatively hereafter. This concentration corresponds to approximately 2500 molecules for a cell of diameter 20 μm. We notice that in most of the proapoptotic protein deficiencies, the model predicts malfunction in apoptosis, one of the hallmarks of cancer. This is in line with the observation that cancer is linked with suppressed levels of pro-apoptotic proteins.[37] Note that when the production rates of proapoptotic proteins are reduced to 50 % of their nominal values, cells with malfunction in apoptosis were predicted to be unattainable in all cases for parameter sets 1 and 2 and for parameter set 3 except for Bid deficiency, the cells had malfunction in apoptosis (data not shown).

The final value of the caspase-3 concentration was obtained as the steady-state solution of the governing differential equations using XPPAUT.[31] At the end of the simulation run, if it was found that the cell is resistant to apoptosis, then we simulated the effect of the potential treatments described below. To this end, we made an appropriate change in a parameter to represent the effect of treatment and checked whether caspase-3 was produced. We assumed that the treatment becomes successful if caspase-3 concentration reached values greater than or equal to 1 nM. In those cells apoptosis may occur in a monostable fashion as well. This was not investigated in this work.

Using the sensitivity results (Table 1), four different treatment methods, which were also proposed by experimentalists, and which may yield the final caspase-3 concentration greater than or equal to 1 nM were simulated to achieve apoptosis in the hypothetical malfunctioning cell:

1) Proteasome inhibitor treatment[38] (simulated by reducing the degradation rate constant ($k_d$) of proteins)

2) Bcl-2 inhibitor treatment[39] (simulated by decreasing the production rate constant of anti-apoptotic protein Bcl-2)

3) IAP inhibitor treatment[39] (simulated by decreasing the production rate constant of anti-apoptotic protein IAP)

4) Bid-like synthetic peptides treatment[40] (simulated by increasing the production rate of proapoptotic protein Bid)

The first treatment listed above which has the effect of reducing the degradation rate constants of the proteins might also have severe side effects because this treatment inhibits proteasomes that degrade proteins involved in pathways other than apoptosis as well and hence affecting the other functions of the cell. On the other hand, downregulating Bcl-2 and IAP and upregulating Bid might have less severe side effects.

The predictions of the outcomes of those four potential treatments are discussed below and the results are presented in Tables 3-9.

**Possible outcomes of proteasome inhibitor treatment.**

Herein we checked whether apoptosis is achieved in a hypothetical malfunctioning cell, i.e., a cell whose steady-state concentration of caspase-3 is less than 1 nM. For this purpose, as a treatment, we reduced the degradation rate constant of proteins ($k_d$) to achieve apoptosis. Rate constant $k_d$ can be reduced biochemically by using a proteasome inhibitor.[38] The preclinical studies have shown that proteasome inhibitor bortezomib induces apoptosis, and overcomes chemoresistance in several malignancy models *in vitro* and *in vivo*.[41] As the proteins are degraded by the same proteasome machinery, we have taken all the degradation rates of the proteins to be equal to $k_d$ (0.006 s$^{-1}$ as an approximate value) and decreased its value to see if caspase-3 concentration exceeds the threshold value of 1 nM to mimic the effect of proteasome inhibitor. When the cell is a hypothetical malfunctioning cell, then, the ranges of $k_d$ values within which apoptosis is possible are tabulated in Table 3. It is shown that as $k_d$ value becomes smaller, the steady-state caspase-3 concentration increases in the present theoretical results (sensitivity results in Table 1). This functional relation was investigated using steady-state concentration of caspase-3 versus $k_d$ graph. This bifurcation diagram of caspase-3 with respect to $k_d$ is presented in Figure 2. The parameter set 1 was used for the values of the parameters in the rate equations except for the Apaf-1 production rate which was decreased by a factor of 100-fold (upper left entry in Table 3) to create a hypothetical malfunctioning cell. Apoptotic response was monostable apoptosis when $k_d$ was less than the limit point 1 (LP1) ($k_d$ = 0.006x0.07 s$^{-1}$) or monostable cell survival when $k_d$ was greater than limit point 2 (LP2) ($k_d$ = 0.006x0.48 s$^{-1}$) and bistable when $k_d$ was in between this interval (0.006 s$^{-1}$ is the nominal value of $k_d$). For the bistable case, whether the response results in cell survival or apoptosis depends on the initial concentration of caspase-3 protein. The upper solid curve in Figure 2 represents the apoptotic steady state, the dashed curve in the inset represents the unstable steady state (a slight increase in caspase-3 concentration will lead to apoptosis, whereas, a slight decrease will lead to cell survival) and the horizontal lower solid line represents the cell survival steady state. The inset is the enlargement of the lower part of the diagram as this becomes invisible due to scaling. The points for steady states which were simulated are shown in circles and the smooth function curves were obtained using the KaleidaGraph Version 4.0 (Synergy Software). The caspase-3 concentration $6.7 \times 10^{-3}$ nM was very low to start apoptosis when the value of $k_d$ at LP2 is $0.006 \times 0.48$ s$^{-1}$. Therefore, in Table 3, we tabulated the range of $k_d$ values for which caspase-3 steady state concentration is above the threshold value of 1 nM which was obtained when $k_d$ was smaller than or equal to $0.006 \times 0.35$ s$^{-1}$. The same procedure was repeated to fill in the rest of the entries in Table 3. The case when $k_d <$ LP1, i.e., the cells are monostable apoptotic, was not investigated in this work.

In this study, inhibition of proteasomes (simulated by reducing $k_d$) was found to induce apoptosis in hypothetical malfunctioning cells. Experimental studies also suggest that proteasome inhibitors can be used for inducing apoptosis.[38] However, it should be noted that the present study does not take into account the inhibition of proteasomal degradation of IKB and subsequent inhibition of NF-KB which can change the production rates of proteins in the apoptotic pathways.[42]

It can be seen in Table 3 that the overall proapoptotic protein degradation rate has to be decreased in different proportions for each proapoptotic protein deficiency. This difference may help to determine the amount of proteasome inhibitor needed to treat a patient. For a healthy cell, it was assumed that apoptosis is possible when $k_d$ is less than or equal to 0.006 s$^{-1}$.[3] When a cell's Apaf-1 production rate was reduced by 100-fold, while keeping the other protein production rates at their nominal values, apoptosis was then possible for $k_d$ range between $0 - 0.006 \times 0.35$ s$^{-1}$. When a cell's Bid production rate was reduced by the same amount, apoptosis could be achieved if $k_d$ range was $0 - 0.006 \times 0.88$ s$^{-1}$. Therefore, the amount of proteasome inhibitor which should be used for the treatment of a patient with Apaf-1 deficiency may be more than a patient with Bid deficiency according to the present model. To know the least amount of drug that is effective is important to reduce its side effects.

When we analyzed the consequences of Bid deficiency in people whose Bid production rate was reduced to 15 % of its nominal value, it was found that the cells represented by parameter sets 1 and 2 are healthy, however, the cell represented by parameter set 3 have malfunction in apoptosis. Moreover, the $k_d$ ranges are similar in parameter sets 1 and 2 but the ranges are significantly different in parameter set 3. To check whether variations in four proteins' (IAP, procaspase-3, procaspase-9 and mitochondrial cyt *c*) production rates produce statistically significant outcome, we have generated 6 point cells that belong to patients from the parameter sets in Table 4 which are obtained around the nominal one for parameter sets 1 and 3 given in Table 3 (parameter set 2 is not included for it is similar to the parameter set 1) for each hypothetical impairment in the production levels of the pro-apoptotic proteins (column 1 in Table 5) and 90, 95, 99% confidence intervals for the difference in the means of the two hypothetical populations of the parameter sets 1 and 3 are calculated. When the production rates of proapoptotic proteins are decreased by 100 fold, then it is found that there is a statistical significance for Apaf-1, procaspase-3 and procaspase-9 but no statistical significance for Bid, Bax and mitochondrial cyt *c*. This is because the confidence intervals for the former set include zero within their confidence interval hence the mean values can also be equal. On the other hand, when the production rates are decreased by 15%, some of the mutations did not lead to unhealthy cells hence no confidence interval is calculated for that group and for those that we have calculated all the confidence intervals is found to include zero. The groups of cells in parameter set 1 and 3 can be considered as two groups of different genetic background but similar within. The statistical results suggest that parameter set 1 group require different doses of proteasome inhibitor compared to the individuals whose cells can be represented by parameter set 3 if they have a drastic reduction down to 1% in their production rates of Apaf-1 or procaspase-3 or procaspase-9 proteins.

In this study, we did not assess the degradation of proteins by lysosomes as this will simply shift the steady state concentrations of all the proteins in the cell downwards.

**Possible outcomes of Bcl-2 and IAP inhibitors treatments.**

In the previous treatment method, we determined the proapoptotic proteins whose deficiencies may result in hypothetical malfunctioning cell formation. For treatment purposes, we then reduced the antiapoptotic Bcl-2 and IAP production rates to induce apoptosis in these hypothetical malfunctioning cells with proapoptotic protein deficiency from Apaf-1 to cyt *c* (Table 6). The production rate constants of Bcl-2 and IAP can be reduced biochemically by introducing their hypothetical inhibitors into the medium.[39] Such an inhibitor for Bcl-2 is obatoclax mesylate (GX015-070)[43] and for IAP is SMAC peptide derived small compounds.[39] Instead of including the reaction of a protein with an inhibitor, the same effect can also be obtained by reducing the production rate of the same protein. This means that if three proteins are synthesized and one of them is quickly inactivated by an inhibitor which binds irreversibly, then the production rate of the functional protein will be reduced by one third.

The outcomes of treatment methods for the hypothetical malfunctioning cells wherein Bcl-2 and IAP inhibitors are introduced as a treatment are summarized in Tables 6 and 7, respectively. As for the cells whose caspase-3 concentration exceeds the threshold value of 1 nM, no treatment is needed (denoted by "Healthy cell" in the Tables). On the other hand, when the cells are hypothetical malfunctioning cells then whether the treatment induces apoptosis (the cells are sensitive or resistant to treatment) depends on the parameter set and proapoptotic protein deficiency. Reducing the production rates of Bcl-2 or IAP may induce apoptosis. By how much amount the production rate had to be reduced is presented in Table 6 so that caspase-3 concentration exceeded its threshold value (however IAP inhibitor treatment did not induce apoptosis). For example, in Table 6 for parameter set 1, when Bax concentration was reduced to 15 % of its nominal value, apoptosis was possible if the production rate of Bcl-2 was reduced to values smaller than $3 \times 10^{-2} \times 0.10$ nM/s (note that for a healthy person Bcl-2 formation rate was assumed to be equal to $3 \times 10^{-2}$ nM/s - reference[3]). This cell is sensitive to treatment. On the other hand, if the hypothetical malfunctioning cells did not undergo apoptosis even with zero production rates of Bcl-2 and IAP, then this case is presented as "Apoptosis impossible (denoted by x)" in the Tables. For these cases, the steady-state caspase-3 concentration was either zero, or equal to a value smaller than 1 nM and the corresponding hypothetical malfunctioning cells are resistant to treatment. For example, in Table 6 and parameter set 1, when the production rate of procaspase-3 is 15 % of its nominal value, steady state caspase-3 concentration is 0.80 nM.

The results for parameter sets 1 and 2 presented in Tables 6 and 7 show that an individual with a Bax production rate reduced to 15 % of its nominal value could be successfully treated by a Bcl-2 inhibitor but not by an IAP inhibitor. Also, Bcl-2 inhibitor therapy can only be effective on people with Bax deficiency (sensitive to treatment) but not on others (resistant to treatment). The reason for Bcl-2 inhibitor being effective on Bax deficiency is possibly because of the fact that Bcl-2 directly interacts with Bax (Figure S1). These simulation results suggest that molecular heterogeneity in patients can be a reason for different treatment consequences. On the other hand, these qualitative results were not obtained for the parameter set 3. Hence, the model is not robust to the choice of parameter values of $\Omega_{IAP}$, $\Omega_{proc9}$, $\Omega_{proc3}$, $\Omega_{cytcmito}$ in assessing the outcomes of Bcl-2 and IAP inhibitors treatment methods. When Bcl-2 inhibitor treatment is employed for the patients whose proapoptotic protein production rates are reduced to 1% of their nominal values, no variations are detected between the groups of parameter set 1 and 3 and also within the groups of each set (Table 8). However, some differences are observed when the production rates are reduced to 15% of their nominal values for Bax production deficiency.

**Possible outcomes of Bid-like synthetic peptides treatment.**

In order to induce apoptosis in hypothetical malfunctioning cells, we then increased the production rate of Bid in the parameter sets by three-fold. Increase in Bid production rate may be induced biologically by the penetration of Bid-like synthetic peptides into the cells by endocytosis.[40] Apoptosis was not possible in response to Bid-like synthetic peptides treatment for the parameter sets 1, 2 and 3 even though the production rate of Bid was increased by 20-fold (Table 9).

**Results for modified mitochondria-dependent apoptosis model**

In this study, we increased the IAP production rate ($\Omega_{IAP}$) as large as 1000 fold and simulated the caspase-3 concentration. Under these conditions the caspase-3 concentration reached a steady-state value of 1.8 nM, still above the assumed threshold value (1 nM). This is contrary to the expectations since at such a high value of inhibitor production rate, one would expect a very low caspase-3 concentration[8, 17, 32, 44-45]). Therefore, we modified the mitochondria dependent apoptosis model and included the inhibition of caspase-3 by IAP through ubiquitination and subsequent degradation by the proteasome. (See Methods section). The resulting model for parameter set 1 and 2 are again found to be bistable but monostable cell survival for parameter set 3. However, our modified model for this parameter set is bistable when the production rate of mitochondrial cyt *c* is increased by two-fold. Thousand-fold increase in the production rate of IAP did not produce cells with malfunctioning apoptosis in the existing model.

When the IAP degradation mechanism is modified (see Methods section) then, a reasonable increase in the IAP production rate has resulted in monostable cell survival.

**Comparison of theoretical results with experiments.**

A recent experimental study[29] focused on the effect of IAP and Bcl-2 inhibitors in restoring cyt *c* release from mitochondria to cytoplasm and apoptosis in chondrosarcoma cells. We used the modified model to see if we can see this effect *in silico*. To this end, we assigned a nominal production rate value of 0.06 nM/s to IAP and 0.08 nM/s to Bcl-2 which ensured bistability. Then, we increased both production rates by six-fold so as to create hypothetical malfunctioning cells. Later, we simulated the effect of Bcl-2 or IAP inhibitors which resulted in cyt *c* release and caspase-3 activation which were used to treat cell one at a time (Figure 3). The simulation of IAP inhibition is presented in Figure 3A. The IAP production rate is set at its nominal value (corresponds to a level in a healthy cell) whereas Bcl-2 production rate is set at six-fold of its nominal value. The model predicts cyt *c* release to cytoplasm and caspase-3 activation under these conditions in agreement with observations in the reference[29]. The simulation of Bcl-2 is presented in Figure 3B. Again, the model predicts cyt *c* release to cytoplasm and caspase-3 activation in agreement with the study in the reference.[29] Hence, inhibition of either Bcl-2 or IAP is sufficient to restore normal apoptotic function in states where both proteins are constitutively upregulated. Therefore, the computations are in agreement with the observation that changes in more than one protein's levels can play a role in causing malfunctioning of apoptosis.

Di Pietro et al.[30] conducted an extensive study on Genomics, Transcriptomics, Proteomics, Interactomics, Oncogenomics, and Pharmacogenomics of Apoptotic Machinery in *Homo sapiens*. They report mRNA levels of proteins in apoptotic machinery in normal and cancer cells for 13 cancer types. We compared their findings related to transcriptomics of apoptotic machinery with our theoretical predictions. They utilized the data available in Human Transcriptome Map, NCI60 Cancer Microarray Project and Oncogenomics for cancer and normal tissues.[46-48] The authors reported the upregulation and downregulation of gene expression when the gene is up or downregulated by at least three fold in a cancer tissue compared to a normal one. Accordingly, we increased or decreased the protein expression rates that are present in our model by at least three fold in the simulations and checked whether apoptosis can be induced with enough caspase-3 activation. The results are summarized in Table 10 where column number 1 lists the 13 different cancer types, column numbers 2-7 list the change in the levels of mRNA of proteins which are present in our model (extracted from Figure 7A in their paper) and the last column gives our simulation results. In this table 0 denotes no level change, and + and − denote up and downregulation (black, red and green regions, respectively, in their Figure 7A). For example, for ovary cancer, caspase-3 concentration level predicts cell survival (0.9 nM) which corresponds to casp9, Bcl-2, Bid and Apaf-1 expression rates remaining constant and Bax increasing by three fold in cancer cells compared to normal cells and only when caspase-3 expression rate is decreased by 85% (0.15x), do we get cell survival. Among those 13, our theoretical results predict apoptosis malfunctioning in six cancer types. Interestingly, out of seven types that we failed to predict apoptosis malfunction, the three cancer types (pancreas, skin, thyroid cancers) have mutations in BRAF, CDKN2A and TP53 genes (the remaining ten cancers do not have these mutations) that are either not included (BRAF, CDKN2A) or not represented adequately (TP53, data not shown) in our model. These theoretical results suggest that the transcription of these three genes should be included in an apoptosis model to correctly predict the apoptosis malfunction.

To assess the statistical significance of the prediction of deregulated apoptosis in 6 out of 13 cancer types (since the data is taken from real patients who suffer from cancer), we have simulated the $729(=3^6)$ combinations of +, - or 0 (+ obtained by multiplying the production rate of target protein by three, - by dividing by three and zero by leaving unchanged) in the six target genes and found that 435 out of 729 has led to tumor formation. The fact that the ratio of 6/13 is lower than the background rate 435/729, suggests that we might not have included all the necessary proteins into the model which may lead to cell accumulation and this is conjectured in the previous paragraph. The expression dynamics of BRAF, CDKN2A and TP53 genes and the subsequent dynamics of their protein product concentrations are not exactly being understood and hence not included in the model. Had these been included, the prediction ratio could have been as high as 9/13 which is higher than the background rate.

## Conclusion

We have used an ODE model composed of 31 dependent variables obtained from mass-action kinetics with 52 parameters most of which are coming from the kinetics of the reactions. Out of these 52 parameters caspase-3 concentration was found to be sensitive to 10 of them. A subset of these parameters was changed to create 3 healthy cells which are further changed to create hypothetical malfunctioning cells and four different *in silico* treatment methods are used on the hypothetical malfunctioning cells. It is found that the proteasome inhibitor treatment may be the most effective one compared to other treatment methods as this restores apoptosis in cells for all proapoptotic protein deficiencies. On the other hand, in Bcl-2 and IAP inhibitors and Bid upregulation treatment methods only some of the proapoptotic protein deficiencies may be treated. Consequently, depending on the type of the treatment and the identity of the deficient proapoptotic protein, apoptosis may not be induced in a hypothetical malfunctioning cell. It is to be noted that the response to treatments is studied by considering malfunction only in apoptosis but not in other pathways such as multi drug resistance gene pathway. We speculate that the present study is in line with the view that the reason of different outcomes of a chemotherapy method in different people may be their molecular heterogeneity that can be observed through their cancer biomarkers. Our reasoning for this speculation is that a problem occurring in one stage of cancer, i.e., a malfunction in apoptosis, can also be a factor in the overall picture of cancer progression and chemotherapy resistance. Hence, the effect of molecular heterogeneity in apoptosis may also have a role on cancer chemotherapy resistance.

It is argued that biological processes are highly robust to parameter changes.[49] However, the detailed analysis of the existing mitochondria-dependent apoptosis model and its modified version has shown the parameters and the reactions that are more effective than others. Therefore, computational studies like these may be beneficial to help experimentalists to decide which interactions to study and which kinetic parameters to measure.

The results imply that targeted treatments on one protein only i.e., Bcl-2 and IAP inhibitors treatments and Bid-like synthetic peptides treatment are not very effective except when the problem is in the targeted protein. For example, Bcl-2 inhibitor treatment will very likely restore apoptosis in a malfunctioning cell with a defect in its Bcl-2 protein but not on any other proteins. On the other hand, proteasome inhibitor treatment may be much more efficient since this affects all of the proteins in the model. This hypothesis remains to be tested by experiments. Another hypothesis to be tested by experiments which is raised in this study is that a treatment is not likely to be effective if the target protein is not close in the sequence of reactions/interactions in the pathway to the problematic protein. A close interaction between experimentalists and theoreticians may be useful to test the hypothesis arising from computations which will in turn improve the models to generate new hypotheses.[50]

It is to be noted that out of ten parameters that are found to be sensitive to caspase-3 production, only three out of four randomly chosen four parameters are varied. However, this small subset of parameter variations resulted in qualitative differences in all therapy methods and quantitative differences in proteasome inhibitor therapy method. We speculate that although the apoptosis model used can explain healthy and unhealthy functioning of apoptosis, it is insufficient for designing and guiding cancer chemotherapy methods. We also speculate that if the current apoptosis model is further modified, and presently unknown more accurate values of the sensitive parameters are used while also including other hallmarks of cancer then, the resulting model may facilitate the decision of which chemotherapy drug or combinations of drugs to be used when treating patients with known cancer biomarkers. An interesting recent study by Spencer et al.[51] suggested that a significant amount of cell-to-cell variability in TRAIL-induced apoptosis arises from natural differences in protein expression levels hence, not only due to differences in genotype. We plan to compare the contribution of this effect to the contribution of genotypic differences on resistance to treatments in a future computational study.

**Acknowledgments**

We thank Ivet Bahar for insightful comments, Nesrin Ozoren for helpful discussions. We also thank Turkan Haliloglu for providing computational facilities for this study Elife Zerrin Bagci gratefully acknowledges fellowship provided by TÜBİTAK-BİDEB.

34. J.E. Shoemaker and F.J. Doyle, *Biophys. J.*, 2008, **95**, 2610-2623.

35. G.E. Box, W.G. Hunter and J.S. Hunter, *Statistics for Experiments. An Introduction to Design, Data Analysis, and Model Building*, John Wiley and Sons, 1978.

36. V. K. Sharma, A. Sharma, N. Kumar, M. Khandelwal, K.K. Mandapati, S. Horn-Saban, L. Strichman-Almashanu, D. Lancet and S.K. Brahmachari, *BMC Genomics*, 2006, **7**, 258-265.

37. R.A. Weinberg, *The Biology of Cancer*, Garland Science, New York, 2007.

38. D.J. McConkey and K. Zhu, *Drug Resist. Updat.*, 2008, **11**, 164-179.

39. M. Arkin, *Curr. Opin. Chem. Biol.*, 2005, **9**, 317-324.

40. L.D. Walensky, A.L. Kung, I. Escher, T.J. Malia, S. Barbuto, R.D. Wright, G. Wagner, G.L. Verdine, S.J. Korsmeyer, *Science*, 2004, **305**, 1466-1470.

41. H. Ludwig, D. Khayat, C. Giaccone and T. Facon, *Cancer*, 2005, **104**, 1794-1807.

42. V. Poulaki, C.S. Mitsiades, V. Kotoula, J. Negri, D.G. McMillin, J.W. Miller and N. Mitsiades, *Invest. Opthalmol. & Vis. Sci.*, 2007, **48**, 4706-4719.
33

**Figure legends**

Figure 1: Representation of healthy tissues that have homeostasis in cell number and unhealthy tissues that have cell accumulation. A. Phase plane for a mathematical model of apoptosis with suitable values of parameters that ensures bistability in response to apoptotic stimulus. B. Phase plane for an apoptosis model with parameter values that ensure monostable cell survival.

Figure 2: Bifurcation diagram for mitochondria-dependent apoptosis model. The parameter values are those in parameter set 1 except for Apaf-1 production rate is decreased by 100 fold with respect to the nominal value. Inset: Enlargement of the lower portion of the diagram that depicts the limit points clearly.

Figure 3: Model predictions for an apoptosis resistant cell that is treated by IAP and Bcl-2 inhibitors. Model prediction of time evolutions of cyt *c* and caspase-3 concentrations when the apoptosis resistant cell is assumed to be treated by (A) an IAP inhibitor (steady state concentration of caspase-3 is 0.006 μM) (B) a Bcl-2 inhibitor (steady state concentration of caspase-3 is 0.001 μM).

Table 1. Sensitivity analysis of the parameters in mitochondria dependent apoptosis model. Steady-state values of caspase-3 when the parameters are perturbed are presented.

| Parameters | [caspase-3] (nM) | |
|---|---|---|
| | parameter value x 100 | parameter value / 100 |
| $k_d$ | 0 | 4827 |
| $\Omega_{Apaf\text{-}1}$ | 7 | 0 |
| $\Omega_{IAP}$ | 0 | 5.4 |
| $\Omega_{procaspase3}$ | 543 | 0 |
| $\Omega_{procaspase9}$ | 47 | 0 |
| $\Omega_{Bid}$ | 2.2 | 0 |
| $\Omega_{Bcl2}^{o}$ | 0 | 5.5 |
| $\Omega_{Bax}^{o}$ | 5.9 | 0 |
| $\Omega_{cytcmito}$ | 7 | 0 |
| p53 | 5.5 | 0 |

Table 2. The three parameter sets used in simulations of mitochondria-dependent apoptosis model

|  | Parameter set 1 Bagci et al. [3] model | Parameter set 2 | Parameter set 3 |
|---|---|---|---|
| $\Omega_{IAP}$ | $3 \times 10^{-2}$ nM/s | $4.5 \times 10^{-2}$ nM/s | $9 \times 10^{-2}$ nM/s |
| $\Omega_{procaspase3}$ | $3 \times 10^{-1}$ nM/s | $3.6 \times 10^{-1}$ nM/s | $9 \times 10^{-1}$ nM/s |
| $\Omega_{procaspase9}$ | $3 \times 10^{-1}$ nM/s | $2.85 \times 10^{-1}$ nM/s | $3 \times 10^{-1}$ nM/s |
| $\Omega_{cytcmito}$ | $3 \times 10^{-1}$ nM/s | $3 \times 10^{-1}$ nM/s | $1 \times 10^{-1}$ nM/s |

Table 3. Degradation rate constant range in which apoptosis occurs – possible outcomes of proteasome inhibitor treatment

| Deficiency in proapoptotic protein (nominal value multiplied by a factor) | Parameter set 1 | Parameter set 2 | Parameter set 3 |
|---|---|---|---|
| $\Omega_{Apaf-1}$ x 0.01 | 0 – L* x 0.35 s$^{-1}$ | 0 - L x 0.35 s$^{-1}$ | 0 - L x 0.27 s$^{-1}$ |
| $\Omega_{procaspase-3}$ x 0.01 | 0 – L x 0.31 s$^{-1}$ | 0 - L x 0.33 s$^{-1}$ | 0 - L x 0.53 s$^{-1}$ |
| $\Omega_{procaspase-9}$ x 0.01 | 0 – L x 0.42 s$^{-1}$ | 0 - L x 0.43 s$^{-1}$ | 0 - L x 0.50 s$^{-1}$ |
| $\Omega_{Bid}$ x 0.01 | 0 – L x 0.88 s$^{-1}$ | 0 - L x 0.89 s$^{-1}$ | 0 - L x 0.79 s$^{-1}$ |
| $\Omega_{Bax}$ x 0.01 | 0 – L x 0.42 s$^{-1}$ | 0 - L x 0.42 s$^{-1}$ | 0 - L x 0.36 s$^{-1}$ |
| $\Omega_{cytcmito}$ x 0.01 | 0 – L x 0.35 s$^{-1}$ | 0 - L x 0.35 s$^{-1}$ | 0 - L x 0.27 s$^{-1}$ |
| $\Omega_{Apaf-1}$ x 0.15 | 0 – L x 0.78 s$^{-1}$ | 0 - L x 0.79 s$^{-1}$ | 0 - L x 0.62 s$^{-1}$ |
| $\Omega_{procaspase-3}$ x 0.15 | 0 – L x 0.95 s$^{-1}$ | 0 - L x 0.97 s$^{-1}$ | 0 - L x 0.90 s$^{-1}$ |
| $\Omega_{procaspase-9}$ x 0.15 | 0 – L x 0.91 s$^{-1}$ | 0 - L x 0.93 s$^{-1}$ | 0 - L x 0.81 s$^{-1}$ |
| $\Omega_{Bid}$ x 0.15 | Healthy cell | Healthy cell | 0 - L x 0.97 s$^{-1}$ |
| $\Omega_{Bax}$ x 0.15 | 0 – L x 0.94 s$^{-1}$ | 0 - L x 0.94 s$^{-1}$ | 0 - L x 0.81 s$^{-1}$ |
| $\Omega_{cytcmito}$ x 0.15 | 0 – L x 0.78 s$^{-1}$ | 0 - L x 0.79 s$^{-1}$ | 0 - L x 0.62 s$^{-1}$ |

* L = 0.006

Table 4. The parameter sets used in simulations of mitochondria-dependent apoptosis model for statistical evaluation

|  | Parameter set 1 Bagci et al.[3] model → Variation 1A | Parameter set 3 → Variation 3A |
|---|---|---|
| $\Omega_{IAP}$ | $2.85 \times 10^{-2}$ nM/s | $1.002 \times 10^{-1}$ nM/s |
| $\Omega_{procaspase3}$ | $1.65 \times 10^{-1}$ nM/s | $7.77 \times 10^{-1}$ nM/s |
| $\Omega_{procaspase9}$ | $4.59 \times 10^{-1}$ nM/s | $2.92 \times 10^{-1}$ nM/s |
| $\Omega_{cytcmito}$ | $3.54 \times 10^{-1}$ nM/s | $1.90 \times 10^{-1}$ nM/s |
|  | Parameter set 1 Bagci et al.[3] model → Variation 1B | Parameter set 3 → Variation 3B |
| $\Omega_{IAP}$ | $2.803 \times 10^{-2}$ nM/s | $8.39 \times 10^{-2}$ nM/s |
| $\Omega_{procaspase3}$ | $1.42 \times 10^{-1}$ nM/s | $9.13 \times 10^{-1}$ nM/s |
| $\Omega_{procaspase9}$ | $3.29 \times 10^{-1}$ nM/s | $2.32 \times 10^{-1}$ nM/s |
| $\Omega_{cytcmito}$ | $4.54 \times 10^{-1}$ nM/s | $1.06 \times 10^{-1}$ nM/s |
|  | Parameter set 1 Bagci et al.[3] model → Variation 1C | Parameter set 3 → Variation 3C |
| $\Omega_{IAP}$ | $3.82 \times 10^{-2}$ nM/s | $9.33 \times 10^{-2}$ nM/s |
| $\Omega_{procaspase3}$ | $2.98 \times 10^{-1}$ nM/s | $8.86 \times 10^{-1}$ nM/s |
| $\Omega_{procaspase9}$ | $3.17 \times 10^{-1}$ nM/s | $3.29 \times 10^{-1}$ nM/s |
| $\Omega_{cytcmito}$ | $3.11 \times 10^{-1}$ nM/s | $2.25 \times 10^{-1}$ nM/s |
|  | Parameter set 1 Bagci et al.[3] model → Variation 1D | Parameter set 3 → Variation 3D |
| $\Omega_{IAP}$ | $1.66 \times 10^{-2}$ nM/s | $8.81 \times 10^{-2}$ nM/s |
| $\Omega_{procaspase3}$ | $1.41 \times 10^{-1}$ nM/s | $1.07$ nM/s |
| $\Omega_{procaspase9}$ | $3.71 \times 10^{-1}$ nM/s | $2.29 \times 10^{-1}$ nM/s |
| $\Omega_{cytcmito}$ | $2.84 \times 10^{-1}$ nM/s | $2.14 \times 10^{-1}$ nM/s |
|  | Parameter set 1 Bagci et al.[3] model → Variation 1E | Parameter set 3 → Variation 3E |
| $\Omega_{IAP}$ | $2.41 \times 10^{-2}$ nM/s | $1.029 \times 10^{-1}$ nM/s |
| $\Omega_{procaspase3}$ | $2.94 \times 10^{-1}$ nM/s | $8.85 \times 10^{-1}$ nM/s |
| $\Omega_{procaspase9}$ | $2.31 \times 10^{-1}$ nM/s | $3.73 \times 10^{-1}$ nM/s |
| $\Omega_{cytcmito}$ | $2.60 \times 10^{-1}$ nM/s | $1.06 \times 10^{-1}$ nM/s |
|  | Parameter set 1 Bagci et al.[3] model → Variation 1F | Parameter set 3 → Variation 3F |
| $\Omega_{IAP}$ | $4.19 \times 10^{-2}$ nM/s | $1.06 \times 10^{-1}$ nM/s |
| $\Omega_{procaspase3}$ | $1.94 \times 10^{-1}$ nM/s | $9.57 \times 10^{-1}$ nM/s |
| $\Omega_{procaspase9}$ | $5.18 \times 10^{-1}$ nM/s | $3.67 \times 10^{-1}$ nM/s |
| $\Omega_{cytcmito}$ | $2.17 \times 10^{-1}$ nM/s | $1.26 \times 10^{-1}$ nM/s |

Table 5. Degradation rate constant range in which apoptosis occurs for different sets of parameters around sets 1 and 3 – possible outcomes of proteasome inhibitor treatment

| Deficiency in proapoptotic protein (nominal value multiplied by a factor) | *Parameter set 1* Variation 1A Variation 1B Variation 1C Variation 1D Varitation 1E Varitation 1F | *Parameter set 3* Variation 3A Variation 3B Variation 3C Variation 3D Variation 3E Variation 3F | Difference in means for 90% confidence interval 95% confidence interval 99% confidence interval |
|---|---|---|---|
| $\Omega_{\text{Apaf-1}}$ x 0.01 | *0.35* <br> 0.36 <br> 0.37 <br> 0.35 <br> 0.33 <br> 0.32 <br> 0.32 | *0.27* <br> 0.32 <br> 0.26 <br> 0.35 <br> 0.33 <br> 0.28 <br> 0.29 | $0.0060 \leq (\mu1-\mu2) \leq 0.0633$ <br> $0.0019 \leq (\mu1-\mu2) \leq 0.0698$ <br> $-0.0082 \leq (\mu1-\mu2) \leq 0.0782$ |
| $\Omega_{\text{procaspase-3}}$ x 0.01 | *0.31* <br> 0.21 <br> 0.19 <br> 0.30 <br> 0.19 <br> 0.28 <br> 0.25 | *0.53* <br> 0.49 <br> 0.49 <br> 0.54 <br> 0.53 <br> 0.55 <br> 0.57 | $0.2471 \leq (\mu1-\mu2) \leq 0.3349$ <br> $0.2370 \leq (\mu1-\mu2) \leq 0.3450$ <br> $0.2240 \leq (\mu1-\mu2) \leq 0.3580$ |
| $\Omega_{\text{procaspase-9}}$ x 0.01 | *0.42* <br> 0.39 <br> 0.33 <br> 0.42 <br> 0.35 <br> 0.37 <br> 0.43 | *0.50* <br> 0.53 <br> 0.48 <br> 0.58 <br> 0.54 <br> 0.53 <br> 0.55 | $0.0687 \leq (\mu1-\mu2) \leq 0.2245$ <br> $0.0508 \leq (\mu1-\mu2) \leq 0.2424$ <br> $0.0277 \leq (\mu1-\mu2) \leq 0.2655$ |
| $\Omega_{\text{Bid}}$ x 0.01 | *0.88* <br> 0.90 <br> 0.90 <br> 0.89 <br> 0.85 <br> 0.84 <br> 0.85 | *0.79* <br> 0.86 <br> 0.77 <br> 0.90 <br> 0.87 <br> 0.80 <br> 0.83 | $-0.0016 \leq (\mu1-\mu2) \leq 0.0684$ <br> $-0.0096 \leq (\mu1-\mu2) \leq 0.0764$ <br> $-0.0199 \leq (\mu1-\mu2) \leq 0.0867$ |
| $\Omega_{\text{Bax}}$ x 0.01 | *0.42* <br> 0.43 <br> 0.43 <br> 0.41 <br> 0.40 <br> 0.39 <br> 0.39 | *0.36* <br> 0.40 <br> 0.35 <br> 0.41 <br> 0.40 <br> 0.37 <br> 0.38 | $-0.0475 \leq (\mu1-\mu2) \leq 0.0946$ <br> $-0.0638 \leq (\mu1-\mu2) \leq 0.1104$ <br> $-0.0848 \leq (\mu1-\mu2) \leq 0.1314$ |
| $\Omega_{\text{cyt}c\text{mito}}$ x 0.01 | *0.35* <br> 0.36 <br> 0.37 <br> 0.35 <br> 0.33 <br> 0.32 <br> 0.32 | *0.27* <br> 0.32 <br> 0.32 <br> 0.35 <br> 0.33 <br> 0.28 <br> 0.29 | $0.0024 \leq (\mu1-\mu2) \leq 0.0510$ <br> $-0.0023 \leq (\mu1-\mu2) \leq 0.0557$ <br> $-0.0103 \leq (\mu1-\mu2) \leq 0.0637$ |
| $\Omega_{\text{Apaf-1}}$ x 0.15 | 0.78<br>0.82<br>0.83<br>0.79<br>0.74<br>0.72<br>0.74 | 0.62<br>0.73<br>0.60<br>0.79<br>0.75<br>0.64<br>0.68 | $0.0114 \leq (\mu_1-\mu_2) \leq 0.1386$<br>$-0.0032 \leq (\mu_1-\mu_2) \leq 0.1532$<br>$-0.0220 \leq (\mu_1-\mu_2) \leq 0.1720$ |
| $\Omega_{\text{procaspase-3}}$ x 0.15 | 0.95<br>0.91<br>0.78<br>0.96<br>0.79<br>0.87<br>0.93 | 0.90<br>Healthy cell<br>0.88<br>Healthy cell<br>Healthy cell<br>0.94<br>0.98 | |
| $\Omega_{\text{procaspase-9}}$ x 0.15 | 0.91<br>0.91<br>0.81<br>0.92<br>0.81<br>0.83<br>0.89 | 0.81<br>0.93<br>0.79<br>Healthy cell<br>0.95<br>0.84<br>0.88 | |
| $\Omega_{\text{Bid}}$ x 0.15 | Healthy cell<br>Healthy cell<br>Healthy cell<br>Healthy cell<br>Healthy cell<br>Healthy cell<br>Healthy cell | 0.97<br>Healthy cell<br>0.95<br>Healthy cell<br>Healthy cell<br>Healthy cell<br>Healthy cell | |
| $\Omega_{\text{Bax}}$ x 0.15 | 0.94<br>0.96<br>0.75<br>0.94<br>0.90<br>0.88<br>0.90 | 0.81<br>0.90<br>0.79<br>0.94<br>0.91<br>0.84<br>0.86 | $-0.0520 \leq (\mu_1-\mu_2) \leq 0.0820$<br>$-0.0674 \leq (\mu_1-\mu_2) \leq 0.0974$<br>$-0.0873 \leq (\mu_1-\mu_2) \leq 0.1173$ |
| $\Omega_{\text{cyt}c\text{mito}}$ x 0.15 | 0.78<br>0.82<br>0.65<br>0.79<br>0.74<br>0.72<br>0.74 | 0.62<br>0.73<br>0.60<br>0.79<br>0.75<br>0.64<br>0.68 | $-0.024 \leq (\mu_1-\mu_2) \leq 0.01140$<br>$-0.0399 \leq (\mu_1-\mu_2) \leq 0.1299$<br>$-0.0603 \leq (\mu_1-\mu_2) \leq 0.1503$ |

* L = 0.006

Table 6. Bcl-2 production rate constant range in which apoptosis occurs and possible other outcomes of Bcl-2 inhibitor treatment

| Deficiency in proapoptotic protein (nominal value multiplied by a factor) | Parameter set 1 | Parameter set 2 | Parameter set 3 |
|---|---|---|---|
| $\Omega_{Apaf\text{-}1}$ x 0.01 | x *<br>[caspase-3] = 0 nM | x<br>[caspase-3] = 0 nM | x<br>[caspase-3] = 0 nM |
| $\Omega_{procaspase\text{-}3}$ x 0.01 | x<br>[caspase-3] = 0 nM | x<br>[caspase-3] = 0 nM | x<br>[caspase-3] = 0 nM |
| $\Omega_{procaspase\text{-}9}$ x 0.01 | x<br>[caspase-3] = 0 nM | x<br>[caspase-3] = 0 nM | x<br>[caspase-3] = 0 nM |
| $\Omega_{Bid}$ x 0.01 | x<br>[caspase-3] = 0 nM | x<br>[caspase-3] = 0 nM | x<br>[caspase-3] = 0 nM |
| $\Omega_{Bax}$ x 0.01 | x<br>[caspase-3] = 0 nM | x<br>[caspase-3] = 0 nM | x<br>[caspase-3] = 0 nM |
| $\Omega_{cyt c mito}$ x 0.01 | x<br>[caspase-3] = 0 nM | x<br>[caspase-3] = 0 nM | x<br>[caspase-3] = 0 nM |
| $\Omega_{Apaf\text{-}1}$ x 0.15 | x<br>[caspase-3] = 0 nM | x<br>[caspase-3] = 0 nM | x<br>[caspase-3] = 0 nM |
| $\Omega_{procaspase\text{-}3}$ x 0.15 | x<br>[caspase-3] = 0.80 nM | x<br>[caspase-3] = 0.92 nM | x<br>[caspase-3] = 0 nM |
| $\Omega_{procaspase\text{-}9}$ x 0.15 | x<br>[caspase-3] = 0.52 nM | x<br>[caspase-3] = 0.59 nM | x<br>[caspase-3] = 0 nM |
| $\Omega_{Bid}$ x 0.15 | Healthy cell | Healthy cell | x<br>[caspase-3] = 0 nM |
| $\Omega_{Bax}$ x 0.15 | $0–3 \times 10^{-2}$ x 0.10 nM/s | $0–3 \times 10^{-2}$ x 0.21 nM/s | x<br>[caspase-3] = 0 nM |
| $\Omega_{cyt c mito}$ x 0.15 | x<br>[caspase-3] = 0 nM | x<br>[caspase-3] = 0 nM | x<br>[caspase-3] = 0 nM |

* x: Apoptosis impossible

Table 7. Possible outcomes of IAP inhibitor treatment

| Deficiency in proapoptotic protein (nominal value multiplied by a factor) | Parameter set 1 | Parameter set 2 | Parameter set 3 |
|---|---|---|---|
| $\Omega_{Apaf-1}$ x 0.01 | x * [caspase-3] = 0 nM | x [caspase-3] = 0 nM | x [caspase-3] = 0 nM |
| $\Omega_{procaspase-3}$ x 0.01 | x [caspase-3] = 0 nM | x [caspase-3] = 0 nM | x [caspase-3] = 0 nM |
| $\Omega_{procaspase-9}$ x 0.01 | x [caspase-3] = 0 nM | x [caspase-3] = 0 nM | x [caspase-3] = 0 nM |
| $\Omega_{Bid}$ x 0.01 | x [caspase-3] = 0 nM | x [caspase-3] = 0 nM | x [caspase-3] = 0 nM |
| $\Omega_{Bax}$ x 0.01 | x [caspase-3] = 0 nM | x [caspase-3] = 0 nM | x [caspase-3] = 0 nM |
| $\Omega_{cytcmito}$ x 0.01 | x [caspase-3] = 0 nM | x [caspase-3] = 0 nM | x [caspase-3] = 0 nM |
| $\Omega_{Apaf-1}$ x 0.15 | x [caspase-3] = 0 nM | x [caspase-3] = 0 nM | x [caspase-3] = 0 nM |
| $\Omega_{procaspase-3}$ x 0.15 | x [caspase-3]=0.79 nM | x [caspase-3]=0.90 nM | x [caspase-3] = 0 nM |
| $\Omega_{procaspase-9}$ x 0.15 | x [caspase-3]=0.48 nM | x [caspase-3]=0.55 nM | x [caspase-3] = 0 nM |
| $\Omega_{Bid}$ x 0.15 | Healthy cell | Healthy cell | x [caspase-3] = 0 nM |
| $\Omega_{Bax}$ x 0.15 | x [caspase-3] = 0 nM | x [caspase-3] = 0 nM | x [caspase-3] = 0 nM |
| $\Omega_{cytcmito}$ x 0.15 | x [caspase-3] = 0 nM | x [caspase-3] = 0 nM | x [caspase-3] = 0 nM |

* x: Apoptosis impossible

Table 8. Bcl-2 production rate constant range in which apoptosis occurs and possible other outcomes of Bcl-2 inhibitor treatment obtained for different sets of parameters around sets 1 and 3.

| Deficiency in proapoptotic protein (nominal value multiplied by a factor) | Parameter set 1<br>Variation 1A<br>Variation 1B<br>Variation 1C<br>Variation 1D<br>Variation 1E<br>Variation 1F | Parameter set 3<br>Variation 3A<br>Variation 3B<br>Variation 3C<br>Variation 3D<br>Variation 3E<br>Variation 3F |
|---|---|---|
| $\Omega_{\text{Apaf-1}}$ x 0.01 | *x * [caspase-3] = 0 nM*<br>x [caspase-3] = 0 nM<br>x [caspase-3] = 0 nM<br>x [caspase-3] = 0 nM<br>x [caspase-3] = 0 nM<br>x [caspase-3] = 0 nM<br>x [caspase-3] = 0 nM | *x [caspase-3] = 0 nM*<br>x [caspase-3] = 0 nM<br>x [caspase-3] = 0 nM<br>x [caspase-3] = 0 nM<br>x [caspase-3] = 0 nM<br>x [caspase-3] = 0 nM<br>x [caspase-3] = 0 nM |
| $\Omega_{\text{procaspase-3}}$ x 0.01 | *x [caspase-3] = 0 nM*<br>x [caspase-3] = 0 nM<br>x [caspase-3] = 0 nM<br>x [caspase-3] = 0 nM<br>x [caspase-3] = 0 nM<br>x [caspase-3] = 0 nM<br>x [caspase-3] = 0 nM | *x [caspase-3] = 0 nM*<br>x [caspase-3] = 0 nM<br>x [caspase-3] = 0 nM<br>x [caspase-3] = 0 nM<br>x [caspase-3] = 0 nM<br>x [caspase-3] = 0 nM<br>x [caspase-3] = 0 nM |
| $\Omega_{\text{procaspase-9}}$ x 0.01 | *x [caspase-3] = 0 nM*<br>x [caspase-3] = 0 nM<br>x [caspase-3] = 0 nM<br>x [caspase-3] = 0 nM<br>x [caspase-3] = 0 nM<br>x [caspase-3] = 0 nM<br>x [caspase-3] = 0 nM | *x [caspase-3] = 0 nM*<br>x [caspase-3] = 0 nM<br>x [caspase-3] = 0 nM<br>x [caspase-3] = 0 nM<br>x [caspase-3] = 0 nM<br>x [caspase-3] = 0 nM<br>x [caspase-3] = 0 nM |
| $\Omega_{\text{Bid}}$ x 0.01 | *x [caspase-3] = 0 nM*<br>x [caspase-3] = 0 nM<br>x [caspase-3] = 0 nM<br>x [caspase-3] = 0 nM<br>x [caspase-3] = 0 nM<br>x [caspase-3] = 0 nM<br>x [caspase-3] = 0 nM | *x [caspase-3] = 0 nM*<br>x [caspase-3] = 0 nM<br>x [caspase-3] = 0 nM<br>x [caspase-3] = 0 nM<br>x [caspase-3] = 0 nM<br>x [caspase-3] = 0 nM<br>x [caspase-3] = 0 nM |
| $\Omega_{\text{Bax}}$ x 0.01 | *x [caspase-3] = 0 nM*<br>x [caspase-3] = 0 nM<br>x [caspase-3] = 0 nM<br>x [caspase-3] = 0 nM<br>x [caspase-3] = 0 nM<br>x [caspase-3] = 0 nM<br>x [caspase-3] = 0 nM | *x [caspase-3] = 0 nM*<br>x [caspase-3] = 0 nM<br>x [caspase-3] = 0 nM<br>x [caspase-3] = 0 nM<br>x [caspase-3] = 0 nM<br>x [caspase-3] = 0 nM<br>x [caspase-3] = 0 nM |
| $\Omega_{\text{cyt}c\text{mito}}$ x 0.01 | *x [caspase-3] = 0 nM*<br>x [caspase-3] = 0 nM<br>x [caspase-3] = 0 nM<br>x [caspase-3] = 0 nM<br>x [caspase-3] = 0 nM<br>x [caspase-3] = 0 nM<br>x [caspase-3] = 0 nM | *x [caspase-3] = 0 nM*<br>x [caspase-3] = 0 nM<br>x [caspase-3] = 0 nM<br>x [caspase-3] = 0 nM<br>x [caspase-3] = 0 nM<br>x [caspase-3] = 0 nM<br>x [caspase-3] = 0 nM |
| $\Omega_{Apaf-1}$ x 0.15 | *x [caspase-3] = 0 nM*<br>x [caspase-3] = 0 nM<br>x [caspase-3] = 0 nM<br>x [caspase-3] = 0 nM<br>x [caspase-3] = 0 nM<br>x [caspase-3] = 0 nM<br>x [caspase-3] = 0 nM | *x [caspase-3] = 0 nM*<br>x [caspase-3] = 0 nM<br>x [caspase-3] = 0 nM<br>x [caspase-3] = 0 nM<br>x [caspase-3] = 0 nM<br>x [caspase-3] = 0 nM<br>x [caspase-3] = 0 nM |
| $\Omega_{procaspase-3}$ x 0.15 | *x [caspase-3] = 0.80nM*<br>x [caspase-3] = 0 nM<br>x [caspase-3] = 0.48 nM<br>x [caspase-3] = 0.86 nM<br>x [caspase-3] = 0.44 nM<br>x [caspase-3] = 0.52 nM<br>x [caspase-3] = 0.72 nM | *x [caspase-3] = 0 nM*<br>Healthy cell<br>x [caspase-3] = 0 nM<br>Healthy cell<br>Healthy cell<br>x [caspase-3] = 0.48 nM<br>x [caspase-3] = 0.95 nM |
| $\Omega_{procaspase-9}$ x 0.15 | *x [caspase-3] = 0.52nM*<br>x [caspase-3] = 0 nM<br>x [caspase-3] = 0.43 nM<br>x [caspase-3] = 0.60 nM<br>x [caspase-3] = 0.26 nM<br>x [caspase-3] = 0.18 nM<br>x [caspase-3] = 0.37 nM | *x [caspase-3] = 0 nM*<br>x [caspase-3] = 0 nM<br>x [caspase-3] = 0 nM<br>Healthy cell<br>x [caspase-3] = 0.59 nM<br>x [caspase-3] = 0 nM<br>x [caspase-3] = 0 nM |
| $\Omega_{Bid}$ x 0.15 | *Healthy cell*<br>Healthy cell<br>Healthy cell<br>Healthy cell<br>Healthy cell<br>Healthy cell<br>Healthy cell | *x [caspase-3] = 0 nM*<br>Healthy cell<br>x [caspase-3] = 0 nM<br>Healthy cell<br>Healthy cell<br>Healthy cell<br>Healthy cell |
| $\Omega_{Bax}$ x 0.15 | *0–3x10$^{-2}$x0.10 nM/s*<br>x [caspase-3] = 0 nM<br>0–3x10$^{-2}$x0.59 nM/s<br>0–3x10$^{-2}$x0.29 nM/s<br>x [caspase-3] = 0.54 nM<br>x [caspase-3] = 0 nM<br>x [caspase-3] = 0.43 nM | *x [caspase-3] = 0 nM*<br>x [caspase-3] = 0 nM<br>x [caspase-3] = 0 nM<br>x [caspase-3] = 0 nM<br>x [caspase-3] = 0.80 nM<br>Healthy cell<br>x [caspase-3] = 0 nM |
| $\Omega_{cytcmito}$ x 0.15 | *x [caspase-3] = 0 nM*<br>x [caspase-3] = 0 nM<br>x [caspase-3] = 0 nM<br>x [caspase-3] = 0 nM<br>x [caspase-3] = 0 nM<br>x [caspase-3] = 0 nM<br>x [caspase-3] = 0 nM | *x [caspase-3] = 0 nM*<br>x [caspase-3] = 0 nM<br>x [caspase-3] = 0 nM<br>x [caspase-3] = 0 nM<br>x [caspase-3] = 0 nM<br>x [caspase-3] = 0 nM<br>x [caspase-3] = 0 nM |

\* x: Apoptosis impossible

Table 9. Possible outcomes of Bid-like synthetic peptides treatment

| Deficiency in proapoptotic protein (nominal value multiplied by a factor) | Parameter group 1 | Parameter group 2 | Parameter group 3 |
|---|---|---|---|
| $\Omega_{Apaf\text{-}1}$ x 0.01 | x *<br>[caspase-3] = 0 nM | x<br>[caspase-3] = 0 nM | X<br>[caspase-3] = 0 nM |
| $\Omega_{procaspase\text{-}3}$ x 0.01 | x<br>[caspase-3] = 0 nM | x<br>[caspase-3] = 0 nM | X<br>[caspase-3] = 0 nM |
| $\Omega_{procaspase\text{-}9}$ x 0.01 | x<br>[caspase-3] = 0 nM | x<br>[caspase-3] = 0 nM | X<br>[caspase-3] = 0 nM |
| $\Omega_{Bax}$ x 0.01 | x<br>[caspase-3] = 0 nM | x<br>[caspase-3] = 0 nM | X<br>[caspase-3] = 0 nM |
| $\Omega_{cytcmito}$ x 0.01 | x<br>[caspase-3] = 0 nM | x<br>[caspase-3] = 0 nM | X<br>[caspase-3] = 0 nM |
| $\Omega_{Apaf\text{-}1}$ x 0.15 | x<br>[caspase-3] = 0 nM | x<br>[caspase-3] = 0 nM | X<br>[caspase-3] = 0 nM |
| $\Omega_{procaspase\text{-}3}$ x 0.15 | x<br>[caspase-3] = 0.78nM | x<br>[caspase-3] = 0.88nM | X<br>[caspase-3] = 0.32nM |
| $\Omega_{procaspase\text{-}9}$ x 0.15 | x<br>[caspase-3] = 0.53nM | x<br>[caspase-3] = 0.59nM | X<br>[caspase-3] = 0 nM |
| $\Omega_{Bax}$ x 0.15 | x<br>[caspase-3] = 0.21nM | x<br>[caspase-3] = 0.22nM | X<br>[caspase-3] = 0 nM |
| $\Omega_{cytcmito}$ x 0.15 | x<br>[caspase-3] = 0 nM | x<br>[caspase-3] = 0 nM | X<br>[caspase-3] = 0 nM |

* x: Apoptosis impossible

Table 10. Comparison of the results from Figure 7A in paper by di Pietro et al.[30] and the present simulation results.

|  | Casp3 | Casp9 | Bax | Bcl-2 | Bid | Apaf-1 | Theoretical Apoptotic response |
|---|---|---|---|---|---|---|---|
| Leukaemia | 0 | 0 | + | + | - | - | Monostable cell survival |
| neuroblastoma | 0 | - | + | + | + | - | Monostable cell survival |
| Breast | - | 0 | 0 | + | 0 | 0 | Monostable cell survival |
| Colon | 0 | 0 | 0 | 0 | 0 | 0 | [caspase-3]=5.4 nM |
| Ovary | (x0.15) | 0 | + | 0 | 0 | 0 | [caspase-3]=0.9 nM |
| Kidney | 0 | + | 0 | - | 0 | 0 | [caspase-3]=14.7 nM |
| Skin | 0 | 0 | 0 | 0 | + | 0 | [caspase-3]=5.4 nM |
| Prostate | - | 0 | 0 | 0 | 0 | 0 | Monostable cell survival |
| Pancreas | 0 | 0 | + | + | + | + | [caspase-3]=20.7 nM |
| Stomach | + | 0 | + | 0 | 0 | 0 | [caspase-3]=17.3 nM |
| Lung | 0 | + | + | 0 | - | 0 | [caspase-3]=15.1 nM |
| Liver | 0 | 0 | 0 | 0 | - | 0 | Monostable cell survival |
| Thyroid | 0 | 0 | 0 | 0 | 0 | 0 | [caspase-3]=5.4 nM |

Supplementary Table 1. The chemical reactions and physical interactions used in mitochondria-dependent apoptosis model (originally proposed by [1])*.

| Binding-unbinding interactions, catalytic reactions | Reactions of formation (or production) and degradation of proteins |
|---|---|
| casp8 + Bid ↔ casp8_Bid | mRNA → Apaf-1 |
| casp8_Bid → casp8 + tBid | mRNA → IAP |
| tBid → tBid$_{mito}$ | mRNA → procaspase-3 |
| tBid$_{mito}$ + Bax → tBid_Bax$_{mito}$ | mRNA → procaspase-9 |
| tBid.Bax$_{mito}$ + Bax → tBid + (Bax$_{mito}$)$_2$ | mRNA → Bid |
| (Bax$_{mito}$)$_2$ + cytc$_{mito}$ → (Bax$_{mito}$)$_2$ + cytc | mRNA → Bcl-2 |
| Bcl-2 + Bax → Bcl-2.Bax | mRNA → Bax |
| cyt c + Apaf-1 ↔ cytc.Apaf-1 | mRNA → mitochondrial cyt c |
| 7 cytc.Apaf-1 ↔ apop | casp8 → aminoacids |
| apop + procasp9 ↔ apop.procasp9 | Bid → aminoacids |
| Apop.procasp9 + procasp9 ↔ apop.(procasp9)$_2$ | tBid → aminoacids |
| Apop.(procasp9)$_2$ → apop.(casp9)$_2$ | tBid$_{mito}$ → aminoacids |
| Apop.(casp9)$_2$ ↔ apop.casp9 + casp9 | tBid_Bax$_{mito}$ → aminoacids |
| Apop.casp9 ↔ apop + casp9 | Bax → aminoacids |
| casp9 + IAP ↔ casp9.IAP | (Bax$_{mito}$)$_2$ → aminoacids |
| Apop.casp9 + IAP ↔ apop.casp9.IAP | cytc$_{mito}$ → aminoacids |
| Apop.(casp9)$_2$ + IAP ↔ apop.(casp9)$_2$.IAP | cytc → aminoacids |
| procasp3 + casp9 ↔ procasp3.casp9 | Bcl-2 → aminoacids |
| procasp3.casp9 → casp3 + casp9 | Apaf-1 → aminoacids |
| procasp3 + apop.(casp9)$_2$ ↔ procasp3.apop.(casp9)$_2$ | procasp9 → aminoacids |
| procasp3.apop.(casp9)$_2$ → casp3 + apop.(casp9)$_2$ | casp9 → aminoacids |
| casp3 + IAP ↔ casp3.IAP | IAP → aminoacids |
| casp3 + Bid ↔ cap3.Bid | procasp3 → aminoacids |
| cap3.Bid → casp3 + tBid | casp3 → aminoacids |
| casp3 + Bcl-2 ↔ cap3.Bcl-2 | Bcl-2$_{cleaved}$ → aminoacids |
| cap3.Bcl-2 → casp3 + Bcl-2$_{cleaved}$ | |

*p53 increases Bax formation rate and decreases Bcl-2 formation rate [3].

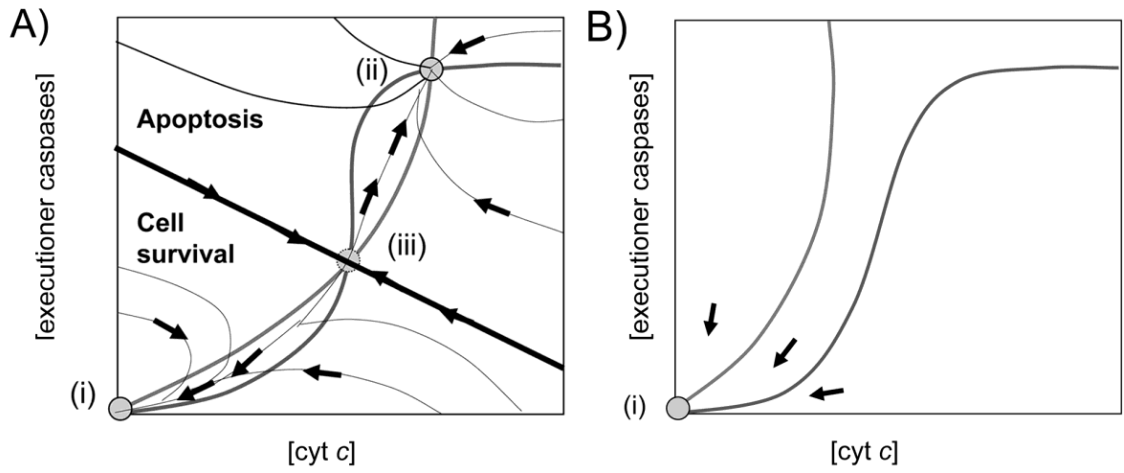

Figure 1

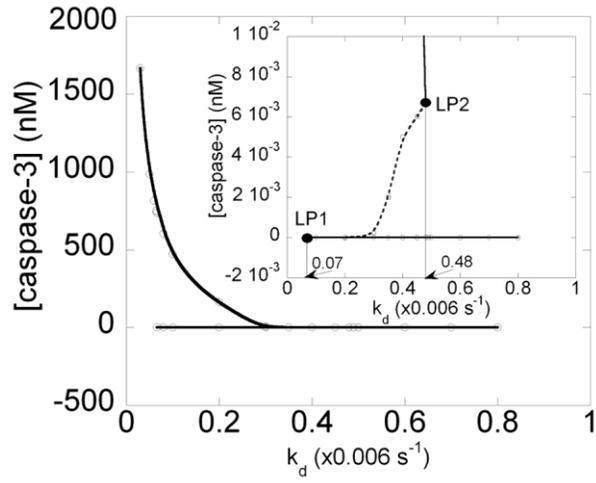

Figure 2

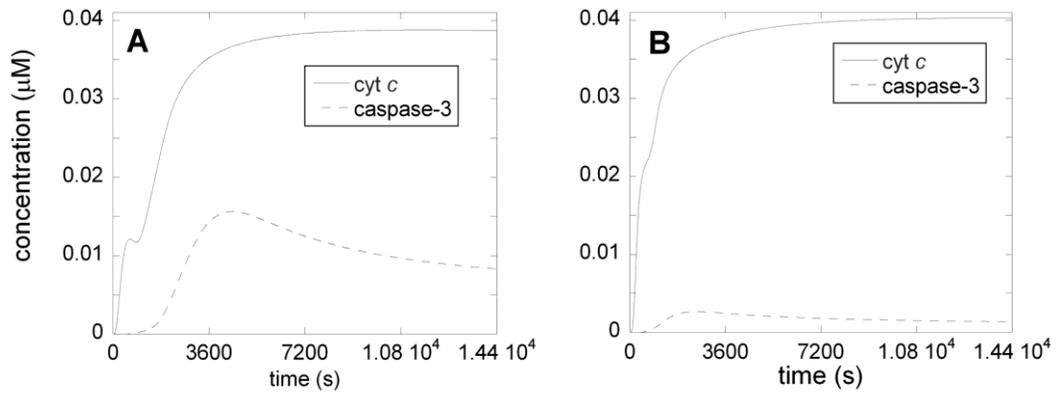

Figure 3